\documentclass[10pt,english,aps,amssymb,prb,twocolumn, showpacs]{revtex4-1}
\pdfoutput=1
\usepackage[T1]{fontenc}
\usepackage[latin9]{inputenc}
\usepackage{amsmath}
\usepackage{graphicx}
\usepackage{amssymb}
\usepackage{esint}
\usepackage{esint}
\usepackage{verbatim}

\makeatletter
\@ifundefined{textcolor}{}
{
 \definecolor{WHITE}{gray}{1}
 \definecolor{RED}{rgb}{1,0,0}
 \definecolor{GREEN}{rgb}{0,1,0}
 \definecolor{BLUE}{rgb}{0,0,1}
 \definecolor{CYAN}{cmyk}{1,0,0,0}
 \definecolor{MAGENTA}{cmyk}{0,1,0,0}
 \definecolor{YELLOW}{cmyk}{0,0,1,0}
 }

\newcommand{\bra}[1]{\langle #1|}
\newcommand{\ket}[1]{|#1\rangle}

\newcommand{\brakets}[2]{\left\langle#1| #2 \right\rangle}
\renewcommand{\phi}{\varphi}
\renewcommand{\epsilon}{\varepsilon}

\renewcommand{\vec}[1]{{\bf #1}}

\usepackage{babel}

\begin{document}

\title {Topological properties of helical Shiba chains with general impurity strength and hybridization}
\author{Alex Weststr\"om}
\author{Kim P\"oyh\"onen}
\author{Teemu Ojanen}
\email[Correspondence to ]{teemuo@boojum.hut.fi}
\affiliation{O. V. Lounasmaa Laboratory (LTL), Aalto University, P.~O.~Box 15100,
FI-00076 AALTO, Finland }
\date{\today}
\begin{abstract}
Recent experiments announced an observation of topological superconductivity and Majorana quasiparticles in Shiba chains, consisting of an array of magnetic atoms deposited on top of a superconductor. In this work we study helical Shiba chains and generalize the microscopic theory of subgap energy bands to a regime where the decoupled magnetic impurity energy and the hybridization of different impurity states can be significant compared to the superconducting gap of the host material. From exact solutions of the Bogoliubov-de Gennes equation we extract expressions for the topological phase boundaries for arbitrary values of the superconducting coherence length. The subgap spectral problem can be formulated as a nonlinear matrix eigenvalue problem from which we obtain an analytical solution for energy bands in the long coherence length limit. Physical consequences and departures from the previously obtained results in the deep-dilute impurity limit are discussed in detail.

\end{abstract}
\maketitle
\bigskip{}

\section{introduction}

Several decades ago it was theoretically predicted that magnetic impurity atoms bind subgap energy states in an ordinary s-wave superconductor.\cite{yu, shiba, rusinov} Since then these Yu-Shiba-Rusinov states, or Shiba states in short, have been studied in detail\cite{salkola}  and observed\cite{yaz} in many Scanning Tunneling Microscopy (STM) experiments. Recent proposals to realize topological superconductivity in chains of magnetic impurities have renewed the interest towards these systems.\cite{choy,np} The attractive features of topological superconductivity in Shiba chains arise from the fact that, in principle, these systems can be realized by garden variety materials; Majorana end states can be directly imaged by STM, and Shiba chains can be made atomically perfect. These extraordinary properties make Shiba chains unique among magnetic \cite{klinovaja2, klinovaja, kjaergaard, ojanen2} and spin-orbit-based\cite{lutchyn,oreg,mourik, das} realizations of topological superconductivity. Recent experimental evidence indicate that Shiba chains indeed support topological phases with accompanying Majorana bound states.\cite{np2} In principle Shiba systems enable probing the non-Abelian statistics of Majorana bound states\cite{li} and could serve as a platform for topological quantum computation.\cite{nayak}           

So far, two distinct mechanisms of topological superconductivity in Shiba chains have been introduced: one relying on helical magnetic order arising from substrate-mediated RKKY interactions in a dilute chain\cite{np,pientka2,pientka3, vazifeh, poyh} and the other arising from the interplay of ferromagnetic order and Rashba spin-orbit interaction on the surface.\cite{li,bry} In the experimental realization, topological superconductivity was observed in ferromagnetic Fe chains deposited on a Pb surface.\cite{np2} In these chains, Fe atoms are in direct contact with each other, which generally leads to ferromagnetic ordering. Both routes to topological superconductivity result in a $p$-wave pairing term in the low-energy theory, providing the link to Kitaev's toy model,\cite{kitaev1} the prototype of 1d topological superconductivity.  However, microscopic theories aiming at a quantitative understanding need to implement the long-range coupling of Shiba states arising from a slow decay of wavefunctions $\frac{e^{-r/\xi}}{r}$ ($\frac{e^{-r/\xi}}{r^{1/2}}$ in 2d) at distances smaller than the superconducting coherence length $\xi$.\cite{pientka2,pientka3,ront,heimes,bry} The long-range nature of the effective tight-binding models lead to significant differences from Kitaev's model and in physically relevant systems the long coherence length limit $\xi\to\infty$ provides an excellent starting point for studies.\cite{pientka2,pientka3}

In this work we study the topological properties of helical Shiba chains. The formation of a magnetic helix from RKKY interaction has been debated recently. In strictly 1d systems the response functions exhibit singular behaviour at twice the Fermi momentum $k=2k_F$  which favours helical ordering of magnetic atoms with the corresponding wave number.\cite{klinovaja3, braun, vazifeh} In higher dimensions the situation is not so clear although numerical evidence supports qualitatively similar behaviour.\cite{reis} However, surface effects and the detailed electronic structure of the host material may give rise to further complications. Following the pioneering paper by Pientka \emph{et al},\cite{pientka2} we study helical chains with arbitrary pitch and tilt angles. The important energy scales in the problem are the single-impurity energy $\epsilon_\alpha=|\Delta|\frac{1-\alpha^2}{1+\alpha^2}$ determined by $\alpha=\pi\nu JS$, where $\nu$ is the density of states, $J$ is the exchange coupling, $S$ is the magnitude of the impurity spin, and $\frac{\alpha |\Delta|}{k_Fa}$ which is the hybridization energy scale of two sites separated by distance $a\ll\xi$. In the treatment of Refs.~\onlinecite{pientka2, pientka3} the topological properties of Shiba chains are solved from an effective long-range tight-binding Hamiltonian. This description is valid for deep impurities close to the Fermi level with energies $\epsilon_\alpha/\Delta\ll 1$ (or $\alpha\approx 1$) in the dilute limit $k_Fa\gg 1$ where the hybridization of different impurity sites is small compared to $\Delta$. In this work we relax these requirements and device a theory valid for when $\epsilon_\alpha$ and $\frac{\Delta}{k_Fa}$ may become significant compared to $\Delta$. Motivation for our work is twofold: on one hand we generalize the theory of subgap bands to new parameter regime which is physically relevant. On the other hand, we can systematically assess the applicability and error of the effective Hamiltonian method in the deep-dilute impurity limit.\\
In Sec.~\ref{model}, we introduce the studied model and show how the Bogoliubov-de Gennes (BdG) equation for a Shiba chain can be formulated as a nonlinear eigenvalue problem for the subgap energy bands and further show how this problem reduces to the effective Hamiltonian description of Ref.~\onlinecite{pientka2} in the deep-dilute impurity limit. In Sec.~\ref{phasediagram} we derive an analytical description of the topological phase diagram as a function of $\alpha$ and $k_Fa$ and analyze the deviations from the deep dilute impurity results of Ref.~\onlinecite{pientka2}. In Sec.~\ref{energybands} we present an analytical solution of the subgap energy bands in the $\xi\to\infty$ limit and compare it to those found in Ref.~\onlinecite{pientka2}. We conclude that the topological properties of the models are practically in perfect agreement for $k_Fa> 10\pi$. In Sec.~\ref{summary} we summarize our findings and discuss the prospects of  treating a ferromagnetic spin-orbit coup!
 led Shiba chains beyond the deep-dilute regime.

\section{microscopic model of helical chains and the nonlinear eigenvalue problem}\label{model}

We consider a number of magnetic impurities on a bulk \textit{s}-wave superconductor. Assuming the impurities are placed at locations $\mathbf{r}_j$, the BdG Hamiltonian describing the system is
\begin{equation}
\mathcal{H} = \left(\frac{k^2}{2m}-\mu\right)\tau_z - J\displaystyle\sum_j \mathbf{S}_j\cdot\mathbf{\sigma}\delta(\mathbf{r}-\mathbf{r}_j) + |\Delta|\tau_x,
\end{equation}
where $k$ and $\mathbf{r}$ denote the momentum and position of the electron, $\Delta$ is the superconducting pairing amplitude, $J$ is the exchange coupling and $\vec{S}_j$ describes the direction and magnitude of the magnetic moment of the $j$th atom.  We will assume that the magnetic ordering of atoms is given by $\hat{\vec{S}}_j=(\cos 2k_haj \sin \theta, \sin 2k_haj \sin \theta, \cos \theta)$, where $k_h$ is the wave number of the magnetic helix pitch angle, $\theta$ is the tilt of the moments and $a$ is the distance between two adjacent moments.  The BdG Hamiltonian is expressed in the Nambu basis  $\hat{\Psi}=(\hat{\psi}_{\uparrow},\hat{\psi}_{\downarrow},\hat{\psi}_{\downarrow}^\dagger,-\hat{\psi}_{\uparrow}^\dagger)^T$ and Pauli matrices $\vec{\tau}$ and $\vec{\sigma}$ describe the particle-hole and the spin degree of freedom. 
In Ref.~\onlinecite{pientka2}  it was shown that the BdG equation $\mathcal{H}\Psi=E\Psi$ leads to the relation  
\begin{equation} \label{TwoBandEnd}
(\mathbf{\hat{S}}_i \cdot \sigma-J_E(0))\Psi(\mathbf{r}_i)=-\sum_{j\neq i}(\mathbf{\hat{S}}_i\cdot \sigma)(\mathbf{\hat{S}}_j\cdot \sigma)J_E(\mathbf{r}_i-\mathbf{r}_j)\Psi(\mathbf{r}_j), 
\end{equation}
where
\begin{widetext}
\begin{equation} \label{J}
J_E(\mathbf{r})=-\frac{\alpha}{\sqrt{|\Delta|^2-E^2}}\frac{e^{-\frac{\sqrt{|\Delta|^2-E^2}}{v_F} r}}{k_Fr}\left[ E\sin k_Fr\,I_{2\times2} +\sqrt{|\Delta|^2-E^2}\cos k_Fr\tau_z+|\Delta|\sin k_Fr\tau_x \right],\quad r>0
\end{equation}
\end{widetext}
and for a vanishing argument 
\begin{equation}
J_E(0)=-\frac{\alpha}{\sqrt{|\Delta|^2-E^2}}\left[ E\,I_{2\times2} +|\Delta|\, \tau_x \right].
\end{equation}
In this expression $\alpha=\pi\nu JS$ as before. The basic assumption in our work is that Eqs.~(\ref{TwoBandEnd}) and~(\ref{J}) provide an accurate description of the physical situation. In contrast to treatments in the deep-dilute limit,\cite{pientka2, pientka3, bry} we do not require that $\frac{1}{k_Fa}\ll 1$. The $1/r$ envelope of the Shiba wavefunctions and $J_E(r)$ are ultimately cut off by microscopic mechanisms sensitive to the precise band structure effects of the substrate metal and details of superconductivity beyond the BCS cutoff scale. However, STM experiments show that the decaying behaviour of Shiba states persists to distances comparable to the Fermi wavelength. Therefore we expect that employing Eq.~(\ref{J}) does not pose a serious restriction.   

We will now proceed to present Eq.~(\ref{TwoBandEnd}) in the basis of decoupled impurity eigenstates.\cite{pientka2} This equates to projecting the spinor onto the basis
\[
\Psi_j =
\begin{pmatrix}
\brakets{+\uparrow}{\Psi_j} &
\brakets{-\downarrow}{\Psi_j}&
\brakets{+\downarrow}{\Psi_j}&
\brakets{-\uparrow}{\Psi_j}
\end{pmatrix}^T,
\]
where $\ket{+\uparrow} = \ket{+}\otimes\ket{\uparrow}$ etc, $|\pm\rangle$ denote eigenstates of $\tau_x$, and $\ket{\Psi_j} = \ket{\Psi(\mathbf{r}_j)}$. We have chosen the order of the spinor elements so that the first two components correspond to the low-energy subspace with decoupled energies $\epsilon_\alpha=\pm|\Delta|\frac{1-\alpha^2}{1+\alpha^2}$. In the transformed basis Eq.~(\ref{TwoBandEnd}) takes the form of a $4N\times4N$ matrix equation:
\begin{widetext}
\begin{equation} \label{ShibaNUM}
\begin{split}
\lambda^2
\begin{pmatrix}
\mathbf{1}+h^{\uparrow\uparrow} & 0 & -h^{\uparrow\downarrow} & 0\\
0 & 0 & 0 & 0\\
-h^{\downarrow\uparrow} & 0 & \mathbf{1}+ h^{\downarrow\downarrow} & 0\\
0 & 0 & 0 & 0\\
\end{pmatrix}
\Psi- \lambda
\begin{pmatrix}
\frac{\mathbf{1}}{\alpha} & D^{\uparrow\downarrow} & 0 & -D^{\uparrow\uparrow}\\
D^{\downarrow\uparrow} & -\frac{\mathbf{1}}{\alpha} & -D^{\downarrow\downarrow} & 0\\
0 & -D^{\downarrow\downarrow} & -\frac{\mathbf{1}}{\alpha} & D^{\downarrow\uparrow}\\
-D^{\uparrow\uparrow} & 0 & D^{\uparrow\downarrow} & \frac{\mathbf{1}}{\alpha}\\
\end{pmatrix}
\Psi-
\begin{pmatrix}
0 & 0 & 0 & 0\\
0 & \mathbf{1}+h^{\downarrow\downarrow} & 0 & -h^{\downarrow\uparrow}\\
0 & 0 & 0 & 0\\
0 & -h^{\uparrow\downarrow} & 0 & \mathbf{1} + h^{\uparrow\uparrow}\\
\end{pmatrix}
\Psi = 0,
\end{split}
\end{equation}
\end{widetext}
In the above we have introduced the $N\times N$ matrices
\begin{equation}
\begin{split}
h^{\sigma\sigma^\prime}_{ij} &\equiv C_{ij}\sin k_Fr_{ij}\bra{\sigma}\sigma^\prime\rangle_{ij}\\ D^{\sigma\sigma^\prime}_{ij} &\equiv C_{ij}\cos k_Fr_{ij}\bra{\sigma}\sigma^\prime\rangle_{ij}.
\end{split}
\end{equation}
and defined
\begin{align}\label{lamb}
&\lambda \equiv \frac{|\Delta|+E}{\sqrt{|\Delta|^2-E^2}},\nonumber\\
&C_{ij} \equiv \frac{e^{-\frac{\sqrt{|\Delta|^2-E^2}}{v_F} r_{ij}}}{k_Fr_{ij}} = \dfrac{\exp(-\frac{r_{ij}}{\xi_E })}{k_Fr_{ij}},
\end{align}
with the prescription $C_{ij} = 0$ for $i = j$. Equation (\ref{ShibaNUM}) should be regarded as a Nonlinear Eigenvalue Problem (NEVP) for $E$ and 4$N$ component eigenspinors $\Psi$.  In deriving Eqs.~(\ref{TwoBandEnd}),(\ref{J}) one has to assume that $|E|<|\Delta|$, so the treatment is valid only for the subgap spectrum. We emphasize that the energy dependence in Eq.~(\ref{ShibaNUM}) only enters through $\lambda$ and $C_{ij}$.  The fact that also the matrix elements - not only $\lambda$ - depend on the energy will generally complicate the treatment considerably. However, we will find efficient techniques to work around this problem. 

Following Ref.~\onlinecite{pientka2}, we perform a unitary transformation after which the spin matrix elements become
\begin{equation}
\begin{split}
\brakets{\uparrow}{\uparrow}_{ij} &= \cos^2\frac{\theta}{2}e^{ik_Hx_{ij}} + \sin^2\frac{\theta}{2}e^{-ik_Hx_{ij}}\\
\brakets{\uparrow}{\downarrow}_{ij} &= \brakets{\downarrow}{\uparrow}_{ij} = i\sin\theta\sin k_Hx_{ij}\\
\brakets{\downarrow}{\downarrow}_{ij} &= \cos^2\frac{\theta}{2}e^{-ik_Hx_{ij}} + \sin^2\frac{\theta}{2}e^{ik_Hx_{ij}}.
\end{split}
\end{equation}
The transform makes the hopping matrix elements translation invariant, enabling us to work in Fourier space. The Fourier transforms
\begin{equation}
h_k^{\sigma\sigma^\prime} = \sum_j h^{\sigma\sigma^\prime}_{ij}e^{ikx_{ij}}
\end{equation}
and $D_k^{\sigma\sigma^\prime}$ can be evaluated as discussed in Appendix A. This transforms the $4N\times4N$ NEVP in Eq.~(\ref{ShibaNUM}) into a compact $4\times4$ form
\begin{widetext}
\begin{equation} \label{ShibaANA}
\begin{split}
\lambda^2
\begin{pmatrix}
1+h^{\uparrow\uparrow}_k & 0 & -h^{\uparrow\downarrow}_k & 0\\
0 & 0 & 0 & 0\\
-h^{\uparrow\downarrow}_k & 0 & 1+ h^{\uparrow\uparrow}_{-k} & 0\\
0 & 0 & 0 & 0\\
\end{pmatrix}
\Psi- \lambda
\begin{pmatrix}
\frac{1}{\alpha} & D^{\uparrow\downarrow}_{k} & 0 & -D^{\uparrow\uparrow}_{k}\\
D^{\uparrow\downarrow}_{k} & -\frac{1}{\alpha} & -D^{\uparrow\uparrow}_{-k} & 0\\
0 & -D^{\uparrow\uparrow}_{-k} & -\frac{1}{\alpha} & D^{\uparrow\downarrow}_k\\
-D^{\uparrow\uparrow}_{k} & 0 & D^{\uparrow\downarrow}_{k} & \frac{1}{\alpha}\\
\end{pmatrix}
\Psi-
\begin{pmatrix}
0 & 0 & 0 & 0\\
0 & 1+h^{\uparrow\uparrow}_{-k} & 0 & -h^{\uparrow\downarrow}_{k}\\
0 & 0 & 0 & 0\\
0 & -h^{\uparrow\downarrow}_{k} & 0 & 1 + h^{\uparrow\uparrow}_{k}\\
\end{pmatrix}
\Psi = 0
\end{split}
\end{equation}
\begin{equation}
\begin{split}
h^{\uparrow\uparrow}_k &= \frac{\cos^2\frac{\theta}{2}}{k_F a}\left[A(k+k_F+k_H) + A(k_F-k-k_H)\right]+\frac{\sin^2\frac{\theta}{2}}{k_F a}\left[A(k+k_F-k_H)+A(k_F-k+k_H)\right]\\
h^{\uparrow\downarrow}_k &= \frac{\sin(\theta)}{2k_Fa}\left[A(k+k_F+k_H) + A(k_F-k-k_H)-A(k_F+k-k_H)-A(k_F-k+k_H)\right]\\
D^{\uparrow\uparrow}_k &= \frac{\cos^2\frac{\theta}{2}}{2k_Fa}\left[f(k_F+k+k_H)+f(k_F-k-k_H)\right]+\frac{\sin^2\frac{\theta}{2}}{2k_Fa}\left[f(k_F+k-k_H)+f(k_F-k+k_H)\right]\\
D^{\uparrow\downarrow}_k &= \frac{\sin\theta}{4k_Fa}\left[f(k+k_F+k_H)-f(k_F+k-k_H)+f(k_F-k-k_H)-f(k_F-k+k_H)\right]\\
h^{\downarrow\downarrow}_k &= h^{\uparrow\uparrow}_{-k},\quad D^{\downarrow\downarrow}_k = D^{\uparrow\uparrow}_{-k},\quad h^{\downarrow\uparrow}_k = h^{\uparrow\downarrow}_k,\quad D^{\downarrow\uparrow}_k = D^{\uparrow\downarrow}_k,
\end{split}
\end{equation}
\end{widetext}
expressed in term of the functions
\begin{align} \label{FTfunctions}
f(k) &\equiv -\ln\left(1+e^{-2\frac{{\sqrt{|\Delta|^2-E^2}}}{v_F}a}-2e^{-\frac{{\sqrt{|\Delta|^2-E^2}}}{v_F}a}\cos ka\right),\nonumber\\
A(k) &\equiv \arctan\left(\frac{\sin(ka)}{e^{\frac{{\sqrt{|\Delta|^2-E^2}}}{v_F}a}-\cos(ka)}\right).
\end{align}

To make the connection to the effective two-band Hamiltonian employed by Pientka and collaborators,\cite{pientka2} we recall that the first two components of $\Psi$ span the low-energy subspace relevant in the deep dilute impurity regime. Therefore the approach of Ref.~\onlinecite{pientka2} can be recovered by considering the upper left $2N\times 2N$ block of Eq.~(\ref{ShibaNUM}) and ignoring its coupling to the lower block, expanding to  linear order in $E$ and considering the case $\alpha\approx 1$. This results in the equation
\begin{equation} \label{2bandx}
E\Psi = \begin{pmatrix}
\epsilon_0 - |\Delta| h_{ij}^{\uparrow\uparrow} & |\Delta| D_{ij}^{\uparrow\downarrow} \\
|\Delta| D_{ij}^{\downarrow\uparrow} & - \epsilon_0 + |\Delta| h_{ij}^{\downarrow\downarrow}
\end{pmatrix}\Psi,
\end{equation}
where $\epsilon_0= |\Delta|(1-\alpha)$ is the single-impurity energy in the deep impurity limit ($\alpha\approx1$).  The matrix on the right-hand side of Eq.~(\ref{2bandx}) coincides with the effective two-band Hamiltonian derived in Ref.~\onlinecite{pientka2}. The effective Hamiltonian in momentum space is given by 
\begin{equation} \label{2bandk}
H_k = \begin{pmatrix}
\epsilon_0 - |\Delta| h_k^{\uparrow\uparrow} & |\Delta| D_k^{\uparrow\downarrow} \\
|\Delta| D_k^{\uparrow\downarrow} & - \epsilon_0 + |\Delta| h_{-k}^{\uparrow\uparrow}
\end{pmatrix}.
\end{equation}
In the following sections we will elucidate the relationship between the full four-band model (\ref{ShibaANA}) and the two-band Hamiltonian description (\ref{2bandk}) and study what happens beyond the deep dilute regime.  

\section{topological phase diagrams}\label{phasediagram}
\begin{figure}

\includegraphics[width=0.99\columnwidth]{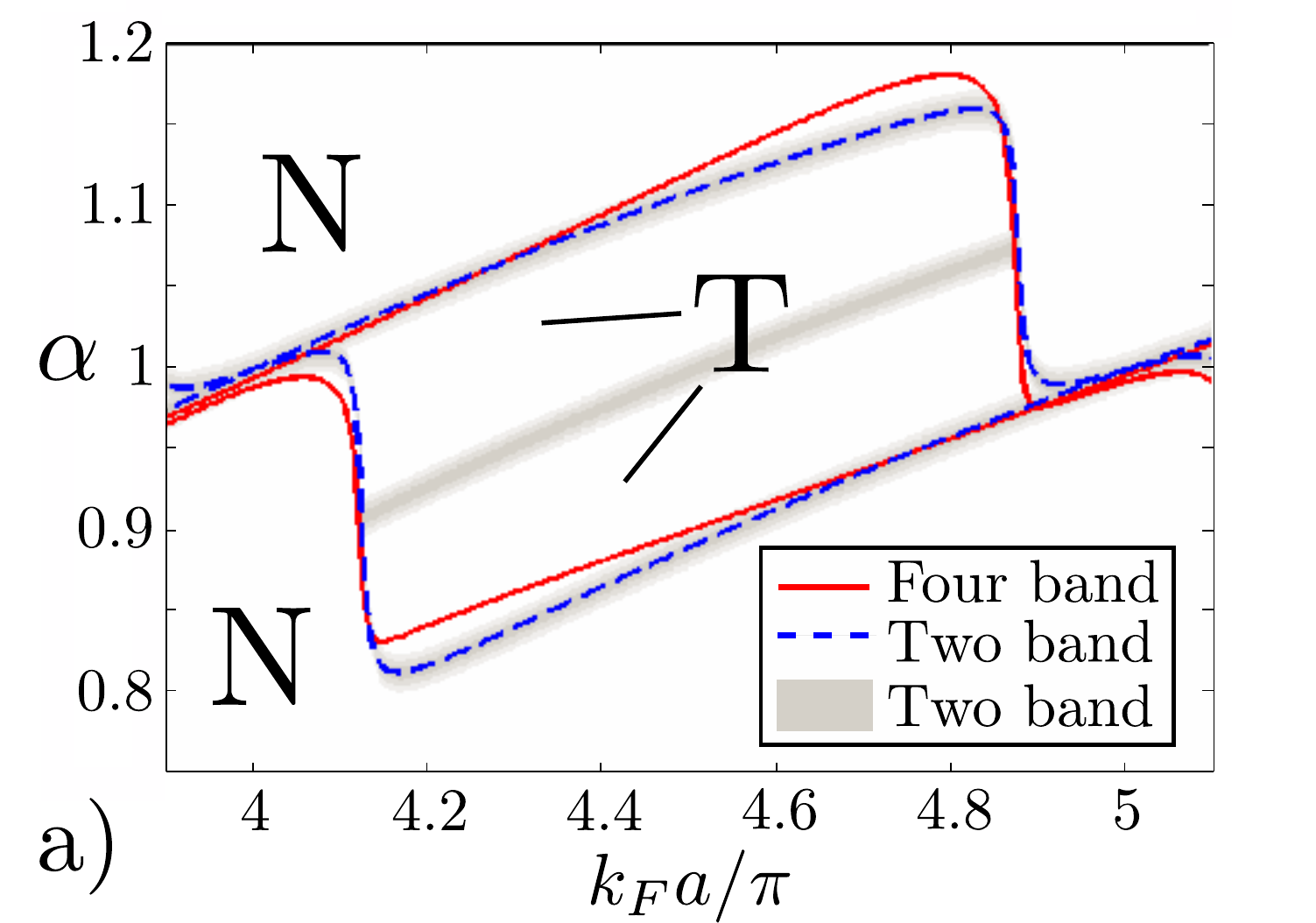}
\includegraphics[width=0.88\columnwidth]{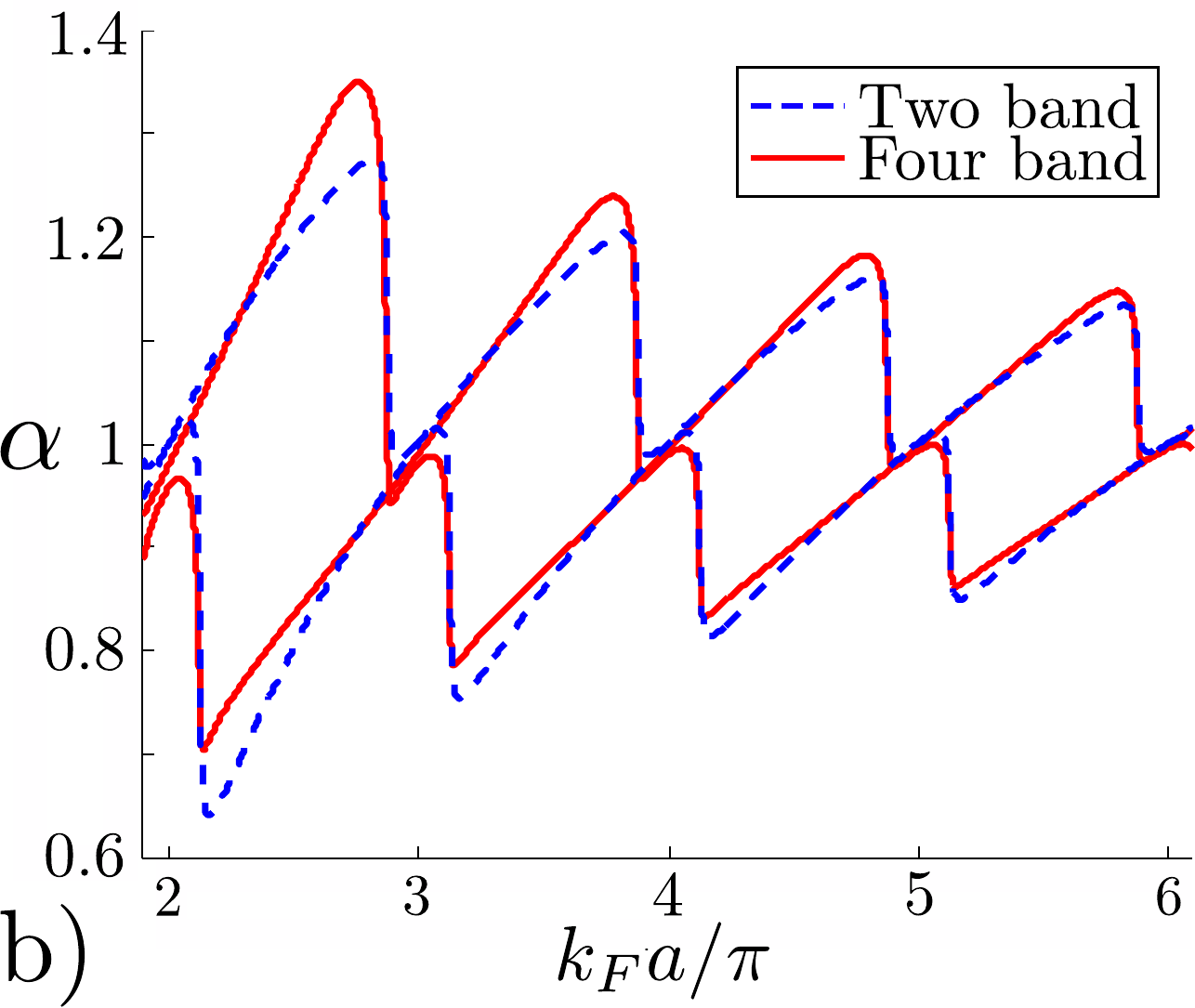}
\caption{(a) Topological phase diagram of a planar helix. Topological and gapped non-topological regions are marked with T and N respectively. The light gray pattern in the background is a numerical calculation of the gap closings based on the two-band model. The dashed line is the topological phase diagram solved analytically as outlined in the text. It is evident that they give the same phase boundaries, but the analytical result does not capture the gap closing in the middle which is not associated with a topological phase transition. The solid red line shows the topological phase diagram for the full four-band model. The parameters used are $k_Ha = \pi/8$, $\theta = \pi/2$, $\xi = 50$, with energies in units of $\Delta$. Note that the parameters are equal to those used in fig.~6(a) in Pientka et al.
(b) A plot of the topological phase diagram over a wider range of values. The topological and non-topological regions are easily identified using (a). We notice that the two models deviate more from each other for low values of $k_Fa$, and as $k_Fa$ increases, the difference between the two models becomes small. The other parameters used are the same as above. }
\label{fig:topo1}
\end{figure}

In the previous section we derived the NEVP describing subgap properties of helical Shiba chains in real space (\ref{ShibaNUM}) and Fourier space (\ref{ShibaANA}). The real space version can be employed in studying properties of finite chains with open boundary conditions while Eq.~(\ref{ShibaANA}) provides a tractable starting point for analytical analysis in an infinite chain. However, in both cases we are faced with a NEVP which poses a considerable complication in terms of solvability. The NEVPs are generalizations of the familiar linear eigenvalue problems (LEVP) - instead of $(\mathbf{A}-\lambda\mathbf{1})\psi = 0$, we have an equation of the form $\mathbf{A}(\lambda)\psi = 0$, where the $N\times N$ matrix $\mathbf{A}$ is a nonlinear function of $\lambda$. As in the case of LEVPs, the eigenvalues of a NEVP can be solved from the noninvertibility requirement $\det(\mathbf{A}(\lambda)) = 0$. For this to be feasible, the $\lambda$-dependence of $\mathbf{A}(\lambda)$ should be relatively simple, for example $\mathbf{A}(\lambda) = f_1(\lambda)\mathbf{A}_1 + f_2(\lambda)\mathbf{A}_{2} + \cdots +f_p(\lambda)\mathbf{A}_p$, where $p\ll N$ in case $N\gg1$. In the case of polynomial eigenvalue problems $\mathbf{A}(\lambda) = \lambda^p\mathbf{A}_p + \lambda^{p-1}\mathbf{A}_{p-1} + \cdots + \mathbf{A}_0$, the NEVP can be transformed to a $pN\times pN$ generalized LEVP by defining new variables $y_p=\psi^p$, so that polynomial NEVPs can be treated by the familiar methods of linear algebra. At first glimpse Eqs.~(\ref{ShibaNUM}) and (\ref{ShibaANA}) appear as polynomial NEVPs for the transformed variable $\lambda=\frac{|\Delta|+E}{\sqrt{|\Delta|^2-E^2}}$; however, the matrix elements also depend on the eigenvalue $E$ through Eqs.~(\ref{lamb}) and (\ref{FTfunctions}). Nevertheless, we can make progress in two important cases: extracting topological phase diagram for arbitrary $\xi_0$ and solving the subgap spectrum in the limit $\xi_0\to\infty$ (the latter will be presented in the next section). 

The solution of the topological phase diagram for planar helix $\theta=\pi/2$ does not require a complete solution of the NEVP (\ref{ShibaANA}) but can be obtained by examining the gap closing at the special points $k=0, \pi/a$ (mod $2\pi$). The logic behind this approach is the following: from the work of Kitaev,\cite{kitaev1} we know that the topological phase for the two-band model (\ref{2bandk}) can be determined by evaluating the invariant $\mathcal{Q}=\mathrm{sign}\left[h(k=0)h(k=\frac{\pi}{a})\right]$, where $h(k)=\epsilon_0-|\Delta| h_k^{\uparrow\uparrow}$. Thus the invariant in the two-band symmetry class (Altland-Zirnbauer class D) can only change when the energy gap closes $E(k)=0$ at $k=0$ and $k=\pi/a$. The topological significance of band-touching points at $k=0, \pi/a$ is shared by the parent four-band model (\ref{ShibaANA}) since topological properties must be shared by the two models.   

We can extract the topological phase boundaries in a form $\alpha(k_Fa)$ from the condition $\det(\mathbf{A}(\lambda)) = 0$, by setting $\lambda = 1\ (E = 0)$ and $k = 0,\pi$. Here $\mathbf{A}(\lambda)$  denotes the three terms in Eq.~ (\ref{ShibaANA}) when written in the form $\mathbf{A}(\lambda)\Psi=0$. This approach has the great advantage that important information can be extracted without solving the full NEVP and the method works for arbitrary coherence lengths $\xi_0$. The only caveat is that this technique does not capture the boundary between the gapless phase present for nonplanar helices $\theta\neq \frac{\pi}{2}$. The gapless phase, arising from a gap closing at generic $k\neq 0, \pi$, is always present for nonplanar helices.\cite{pientka2} However, this approach will yield correct phase boundaries between topological and normal gapped phases even in a nonplanar case if the gapless phase does not overlap with the predicted phase boundary. 

\begin{figure}
\includegraphics[width=0.9\columnwidth]{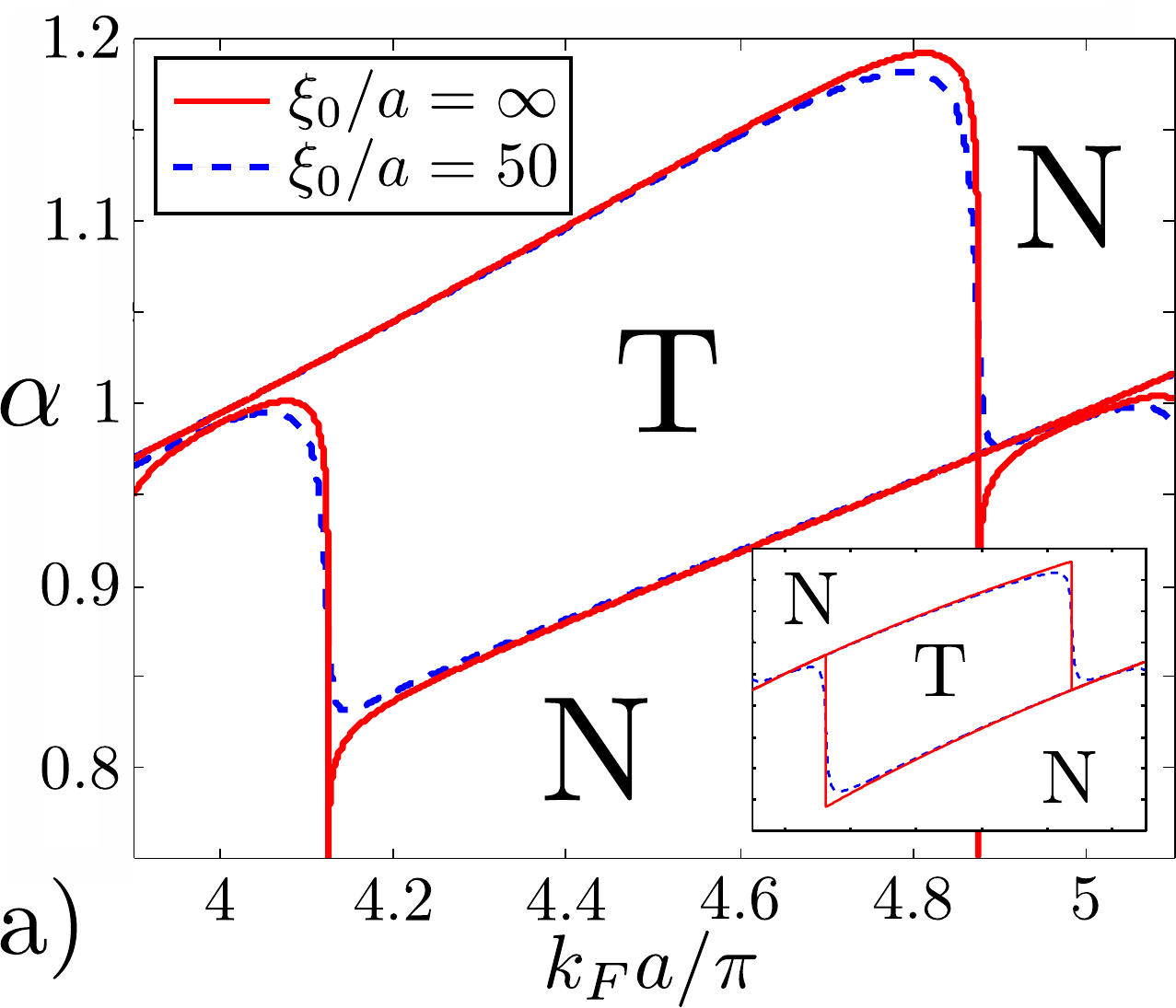}\\
\includegraphics[width=0.9\columnwidth]{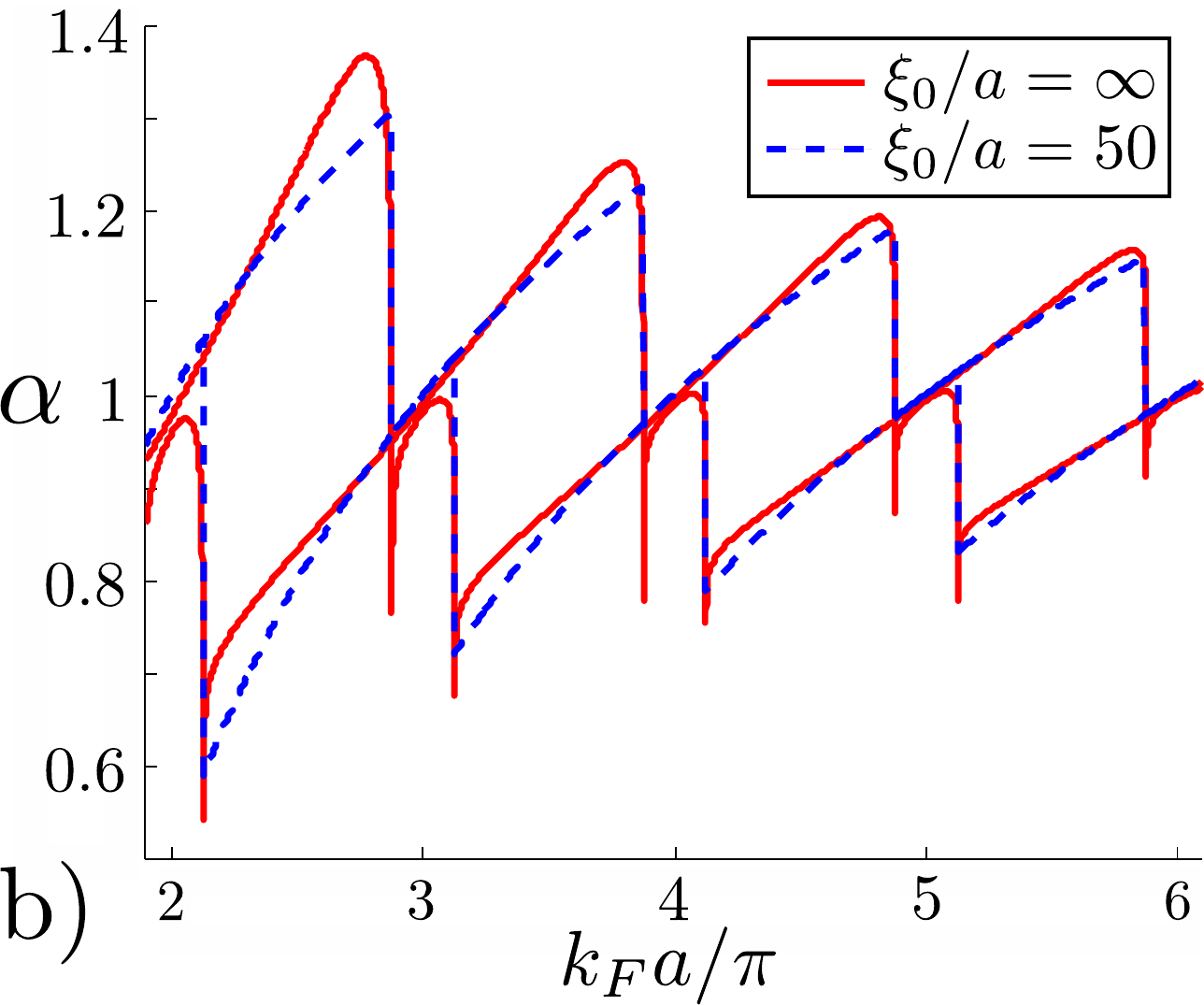} 
\caption{(a) A comparison between the topological diagrams of $\xi_0 = 50$ and $\xi_0 = \infty$ in the the four-band model, with the same $\xi_0$-comparison for the linear two-band model in the inset. (b) A similar comparison as in (a) over a wider range of values. We can infer the topological regions from (a). Both (a) and (b) use $k_Ha = \pi/8$, and energies are in units of $\Delta$. Numerical calculations for $50 < \xi_0 < \infty$ yield lines that lie between those seen in the figures, as would be expected.}
\label{fig:xicomparison}
\end{figure}

We emphasize that the results obtained in this section only fully describe the topological phases in the case $\theta = \pi/2$. The detailed calculation, which proceeds as indicated above, is relegated to Appendix B. The condition for the phase boundaries become
\begin{equation} \label{topodiag4}
\alpha_{0,\pi} = \frac{1}{\sqrt{(1+h_k^{\uparrow\uparrow})^2+(D_k^{\uparrow\uparrow})^2}}\Bigg\vert_{k=0,\pi}
\end{equation}
which is used to plot out the topological phase diagram - note that there is no $\theta$ dependence in this equation. We can compare this result to the equivalent solution for the two-band model in Eq.~(\ref{2bandk}),
\begin{equation} \label{topodiag2}
\alpha_k = 1-\frac{1}{k_Fa}\left[A(k_F-k_H+k) + A(k_F + k_H - k)\right],
\end{equation}
where, as in Eq.~(\ref{topodiag4}), $k$ can either be $0$ or $\pi$. A comparison between the two models is presented in Fig.~\ref{fig:topo1}. As expected, the phase diagram of the full four-band model and effective two-band model are in good agreement in the dilute limit $k_Fa\gg 1$. However, for moderate values $k_Fa\lesssim 4\pi$ the difference of the two models become apparent. Nevertheless, we can conclude that the two-band approximation provides a reasonable description of the topological phases even for a relatively dense impurity chain. Some qualitative differences arise when $\xi_0=\frac{v_F}{|\Delta|}$ increases, as seen in Fig.~\ref{fig:xicomparison}. Most notably, the corners of the topological region in the four-band model either acquire sharp tails or are rounded compared to the two-band approximation.  The qualitative appearance of the phase diagram at high values of $\xi_0\sim 100$ quickly approach that of $\xi_0 \to \infty$.

\section{subgap energy bands in the long coherence length limit } \label{energybands}

The technique introduced in the previous section to extract the phase diagram does not allow for a more detailed description of the system. To investigate the magnitude of the energy gaps and the dispersion of the subgap Shiba bands we need to solve the NEVP in Eq.~(\ref{ShibaANA}). Finding a general solution, even for the $4\times4$ $k$-space NEVP in Eq.~(\ref{ShibaANA}), is a very challenging problem since the energy dependence enters the matrix elements in a very complicated way through the functions $f(k)$ and $A(k)$ in Eq.~(\ref{FTfunctions}). However, considering the long coherence length limit $\xi_E=v_F/\sqrt{|\Delta|^2-E^2}\to \infty$  a crucial simplification takes place. In that case the energy dependence in Eqs.~(\ref{ShibaNUM}) and (\ref{ShibaANA}) \emph{only enters through} $\lambda$. Therefore we can regard  the NEVPs (\ref{ShibaNUM}) and (\ref{ShibaANA}) as quadratic polynomial eigenvalue problems of $\lambda$ in the long coherence length limit. Since $E$ is given straightforwardly by $E=|\Delta|\frac{\lambda^2-1}{\lambda^2+1}$ we can solve the problem by solving the quadratic NEVP for $\lambda$. As discussed in Refs.~\onlinecite{pientka2} and \onlinecite{pientka3}, the long coherence length limit is not a mathematical curiosity but provides an excellent starting point in considering an experimental situation. 

\begin{figure}
\includegraphics[width=0.99\columnwidth]{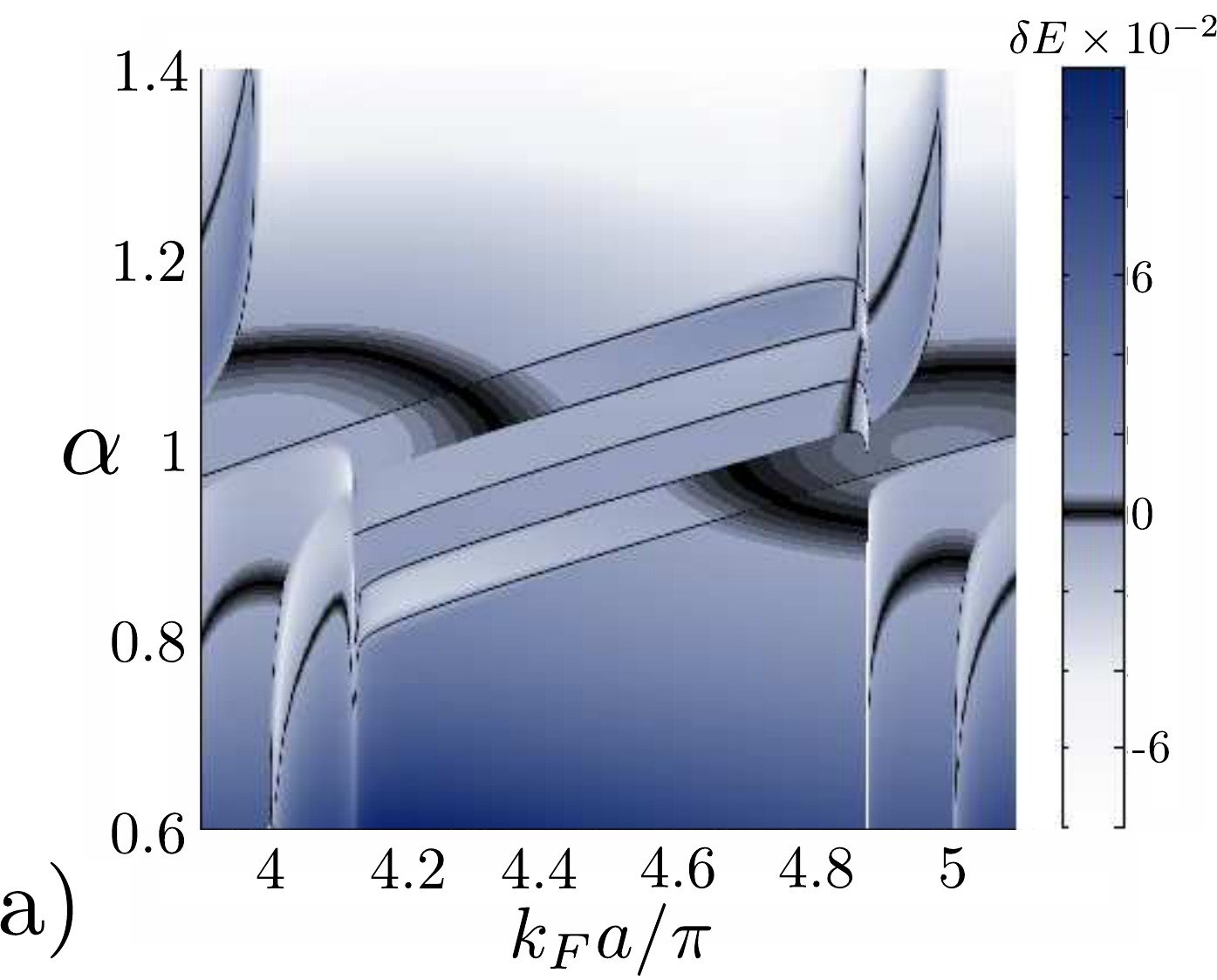}
\includegraphics[width=0.91\columnwidth]{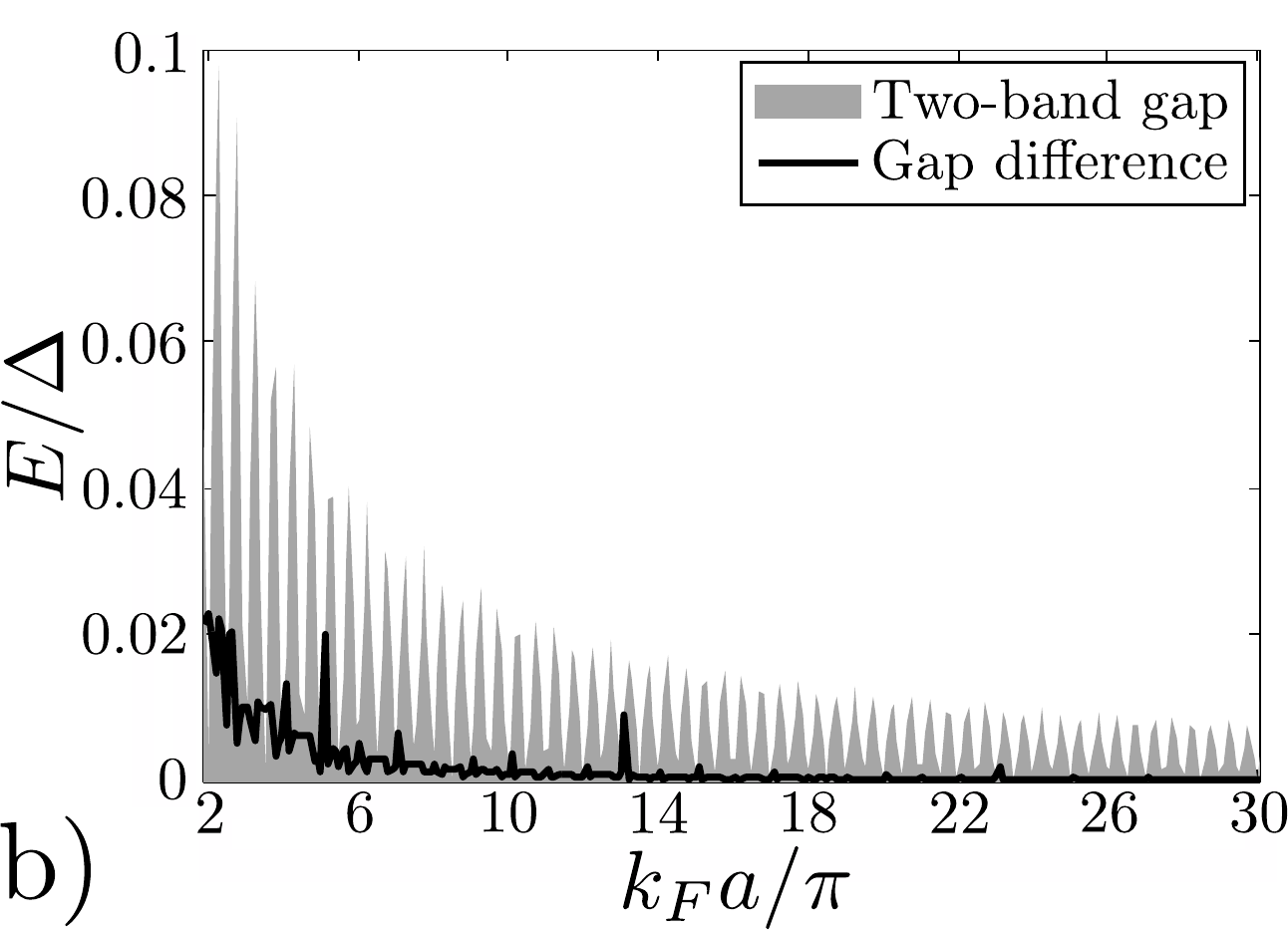}
\caption{ (a) The difference between the minimum gap sizes for the four-band and two-band models. Darker shades generally indicate the four-band gap is larger, while the black areas indicate zones where the gaps are equal. Parameters used are $k_Ha = \pi/8$, $\theta = \pi/2$,  $\xi_0 = \infty$, with energies given in units of $\Delta$.\\
(b) A comparison between the band gap of the two-band model and the absolute difference in minimum gap between the two models. We notice that as $k_Fa$ increases, the difference tends to zero faster than the band gap decreases, as is to be expected. The parameters are the same as previously, except we have fixed $\alpha=1$.}
\label{fig:bandgap1}
\end{figure}

Writing Eq.~(\ref{ShibaANA}) in the form  $\mathbf{A}(\lambda)\Psi = 0$ and evaluating the characteristic polynomial  $\mathrm{det}(\mathbf{A}(\lambda)) = 0$ we obtain an algebraic equation of degree 8 for the eigenvalues $\lambda$. The solution of the this problem is obtained in Appendix C. It turns out, remarkably, that we can find an analytical solution for this problem in relatively compact form given by Eq.~(\ref{app:fullsol}). Even more surprisingly, the solution for the planar helix is given even more compactly by Eq.~(\ref{app:planarsol}). The full energy spectrum is easily calculated from Eqs.~(\ref{app:fullsol}) and (\ref{app:energies}). The found spectrum for the full four-band model can now be compared to the spectrum in the deep dilute case.
From Eq.~(\ref{2bandk}) we find that in the deep-dilute limit the energy of an infinite 1D chain\cite{pientka2} is
\begin{equation}
E_k=|\Delta| \frac{h_{-k}^{\uparrow\uparrow}-h_k^{\uparrow\uparrow}}{2} \pm |\Delta|\sqrt{(D_k^{\uparrow\downarrow})^2+\left(\frac{\epsilon_0}{|\Delta|}-\frac{h_{-k}^{\uparrow\uparrow}+h_k^{\uparrow\uparrow}}{2}\right)^2}. \nonumber
\end{equation}
In Fig.~\ref{fig:bandgap1} we have compared the spectra of the two models. The difference between the minimum gap sizes of the two models is plotted in Fig.~\ref{fig:bandgap1}(a). One clearly sees that in approximately half of the parameter space, the four-band model supports a larger gap than the two-band model. In Fig.~\ref{fig:bandgap1}(b) we see that for small $k_Fa$ the gap difference is significant but vanishes faster than the actual gap size, thus indicating that the two-band model is accurate in the dilute limit. Consequently, our model will agree with Ref.~\onlinecite{pientka2} in the appropriate limit.

\begin{figure}
\includegraphics[width=0.9\columnwidth]{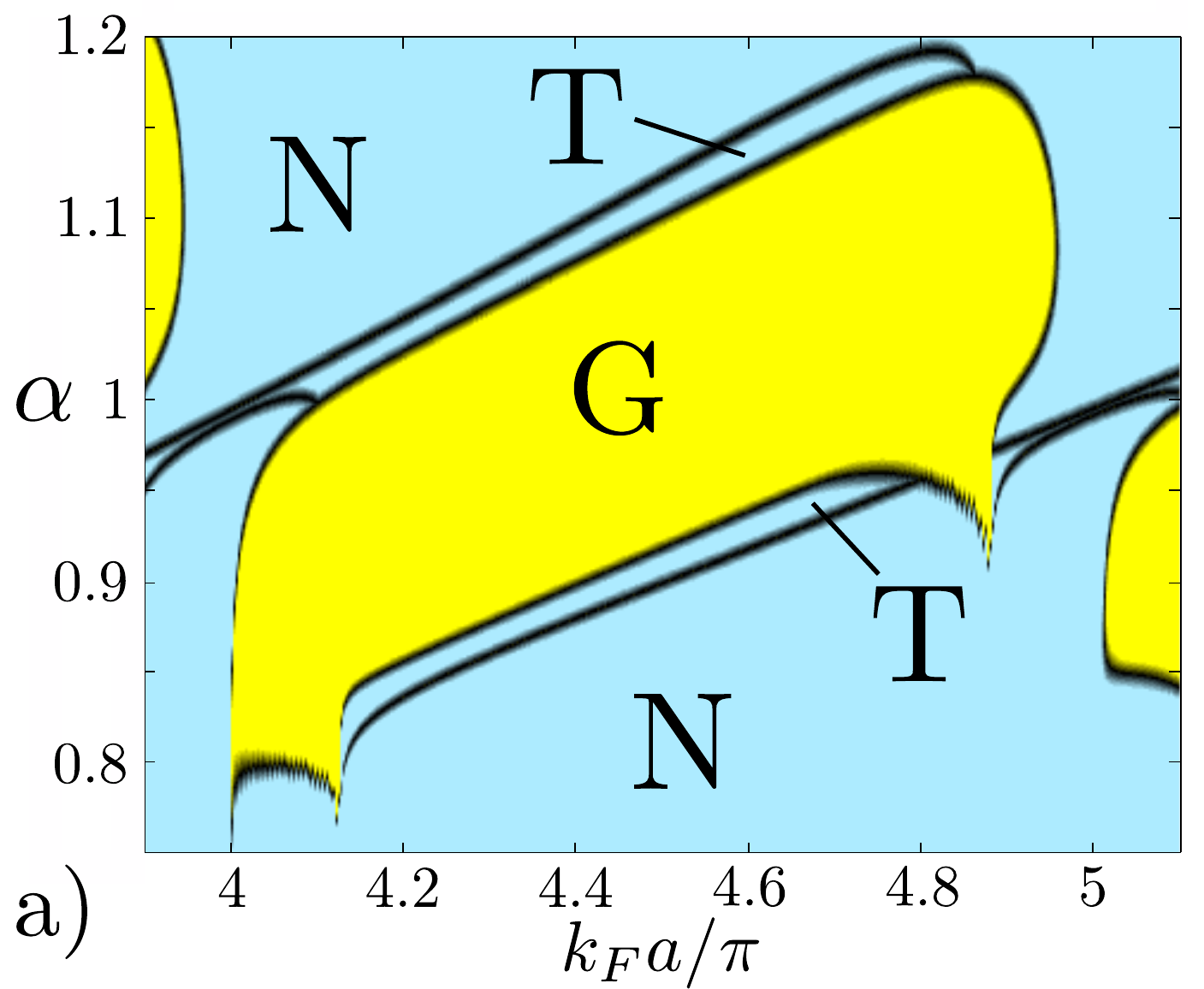} 
\includegraphics[width=0.9\columnwidth]{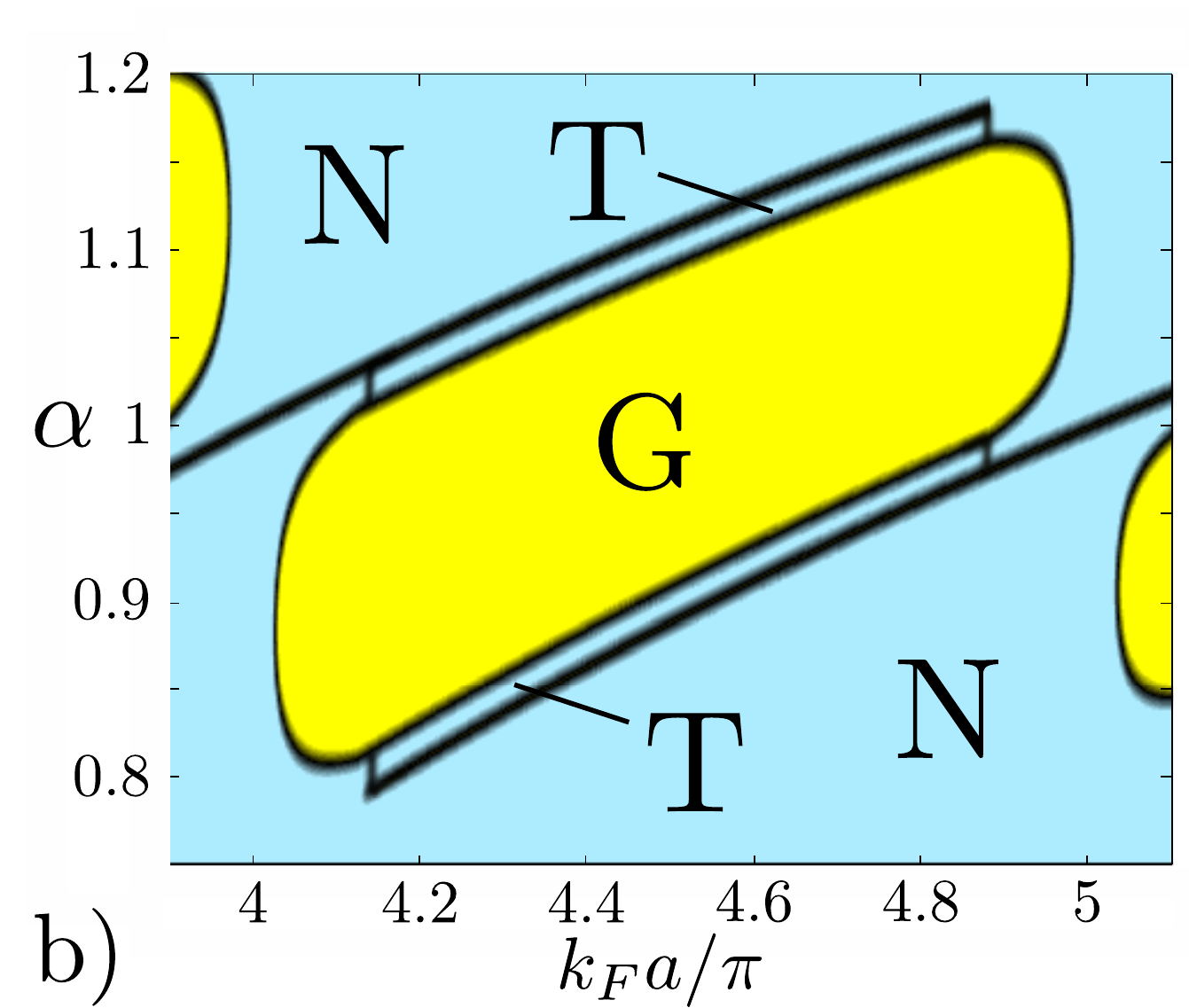}
\caption{(a) Topological phase diagram of a nonplanar helix with $\xi = \infty$ calculated from the spectrum of the four-band model. We have indicated the gapless regions with G and, as before, topological and gapped non-topological regions with T and N. Parameters used $k_Ha = \pi / 8$, $\theta = \pi/5$. (b) The corresponding plot for the two-band model, parameters are the same as in (a). }
\label{fig: thetapiper5}
\end{figure}

The analytical solution of the full problem also allows us to plot study phase diagrams for nonplanar helices $\theta \neq \pi/2$, which we have done in Fig.~\ref{fig: thetapiper5}. As in the two-band model, these values of $\theta$ also give rise to a gapless phase. The true behaviour of the gapless phase have noticable departures from that of the two-band model in some parts of the phase diagram. As seen by comparing Figs.~\ref{fig: thetapiper5} (a) and (b), in the full four-band model the gapless phase occasionally extends to regions that in the deep-dilute limits were gapped, reducing the size of the topological region.

\section{Summary and outlook} \label{summary}

Motivated by recent developments in the pursuit of topological superconductivity in Shiba systems, we generalized the microscopic theory of helical Shiba chains beyond the deep-dilute impurity regime studied in Refs.~\onlinecite{pientka2, pientka3}.  We formulated the Bogoliubov-de Gennes equation for a chain of magnetic impurities as a nonlinear eigenvalue problem which allowed us to solve the topological phase diagram of the system.  We also presented an exact analytical expression for the subgap energy bands in the long coherence length limit. We find that in general, the topological properties of the four-band and the two-band effective Hamiltonian method of Ref.~\onlinecite{pientka2} are in excellent agreement when the parameter $k_Fa$ determining the hybridization of two impurity states separated by $a$ satisfies $k_Fa\gtrsim 10\pi$. Even for smaller values of $k_Fa$ the two-band approximation produces reasonable predictions for topological phase diagrams and energy gaps. The differences in the topological phase diagram between the exact solution and the two-band model become pronounced when $k_Fa\lesssim 4\pi$ and $\xi_E\to\infty$. 

In this work we concentrated on bulk properties of the system. In the topological phase, a finite chain with open boundary conditions supports Majorana end states. Wavefunctions of Majorana end states are generically algebraically decaying in the bulk, but for certain values of the magnetic helix pitch angles they can be essentially exponentially localized, as discovered in Ref.~\onlinecite{pientka3} based on the effective two-band approximation in the deep dilute regime. Going beyond the two-band approximation, Majorana end states can be found by solving the NEVP (\ref{ShibaNUM}) in real space with open boundary conditions. This task is feasible in the long-coherence length limit where the energy-dependence is restricted to $\lambda$ and does not appear in the matrix elements in Eq.~(\ref{ShibaNUM}). We have numerically studied the end states and find that the results of the two-band approximation are in excellent agreement with the exact solution when $k_Fa\gtrsim 10\pi$.

In Ref.~\onlinecite{bry} an effective two-band Hamiltonian was derived for a ferromagnetic chain with a deep-dilute magnetic chain embedded in a 2d electron system with a Rashba spin-orbit coupling. The calculation of Ref.~\onlinecite{bry}  proceeds along the derivation of the two-band model in Ref.~\onlinecite{pientka2} after obtaining the corresponding form of $J_E(r)$ given for a helical model in Eq.~(\ref{J}). A suitable modification of our work could be employed to study the Rashba model of Ref.~\onlinecite{bry} beyond the deep-dilute limit. The Rashba model can also be formulated as a nonlinear eigenvalue problem analogous to Eq.~(\ref{ShibaNUM}). By employing the same change of variables  $\lambda=\frac{|\Delta|+E}{\sqrt{|\Delta|^2-E^2}}$, the corresponding problem can be transformed to a polynomial NEVP in the long coherence length limit. The main difference to Eq.~(\ref{ShibaNUM}) arises from the specific form of the hopping matrix elements that cannot be expressed in terms of elementary functions. Although this complicates analysis, the corresponding NEVP can be treated numerically in real space at least in the long coherence length limit. The ferromagnetic chain is closely related to the experimental realization and the experimental situation seem to reside in the strong hybridization (and long-coherence length) regime  so the dilute limit is not applicable. Therefore modification of our theory to ferromagnetic chains offers an interesting direction of future research.    

\acknowledgments

The authors acknowledge the Academy of Finland for support.

\appendix
\numberwithin{equation}{section}

\section*{Appendix A: Fourier Transforms}
In this appendix we consider the Fourier transforms of the matrices
\begin{equation}
h^{\sigma\sigma^\prime}_{ij} \equiv C_{ij}\sin k_Fr_{ij}\bra{\sigma}\sigma^\prime\rangle\quad D^{\sigma\sigma^\prime}_{ij} \equiv C_{ij}\cos k_Fr_{ij}\bra{\sigma}\sigma^\prime\rangle. \nonumber
\end{equation}
The treatment, again, follows that of Pientka \textit{et al}. Consider, for example, the matrix $h^{\uparrow\downarrow}$. Inserting its expression into the Fourier transform $h(k)^{\sigma\sigma^\prime} = \sum_j h^{\sigma\sigma^\prime}_{ij}e^{ikx_{ij}}$ we find, after some algebraic manipulation,
\begin{widetext}
\[
\begin{split}
h_k^{\uparrow\downarrow} =\frac{\sin\theta}{4k_Fa}\operatorname{Im}\sum_{j=1}^\infty\frac{1}{j}\left[e^{-\frac{aj}{\xi_E}+kaj+k_Haj+k_Faj} -e^{-\frac{aj}{\xi_E}+kaj+k_Haj-k_Faj}-e^{-\frac{aj}{\xi_E}+kaj-k_Haj+k_Faj}+e^{-\frac{aj}{\xi_E}+kaj-k_Haj-k_Faj}\right]\\
-\frac{\sin\theta}{4k_Fa}\operatorname{Im}\sum_{j=1}^\infty\frac{1}{j}\left[e^{-\frac{aj}{\xi_E}-kaj+k_Haj+k_Faj} -e^{-\frac{aj}{\xi_E}-kaj+k_Haj-k_Faj}-e^{-\frac{aj}{\xi_E}-kaj-k_Haj+k_Faj}+e^{-\frac{aj}{\xi_E}-kaj-k_Haj-k_Faj}\right].
\end{split}
\]
\end{widetext}
Recognizing the sums as logarithms, $\ln(1-x) = -\sum_n \frac{x^n}{n}$, and using $\operatorname{Im}(\ln(z)) = \arctan(\operatorname{Im}(z)/\operatorname{Re}(z))$, we obtain the expression used in Eqn.~(11) in the main text. The other matrices are transformed similarly; in some cases, we may have to take the real part instead of the imaginary part, which of course gives $\operatorname{Re}(\ln(z)) = \ln(|z|)$.
\section*{Appendix B: Topological phase diagrams}
We derive the formula for the topological phase diagram as seen in Eqn.~(\ref{topodiag4}), starting from Eqn.~(\ref{ShibaANA}). First we note that at the border between two topological phases, $E = 0$. This corresponds to setting $\lambda = 1$. Topological gap closings occur at $k=0$ or $k = \pm\pi$. We notice that both $D_k^{\uparrow\downarrow}$ and $h_k^{\uparrow\downarrow}$ vanish for these values of $k$, so those terms can be removed. We also notice that $h^{\uparrow\uparrow}_{-\pi} = h^{\uparrow\uparrow}_{\pi}$ and $D^{\uparrow\uparrow}_{-\pi} = D^{\uparrow\uparrow}_{\pi}$, 
further simplifying the problem. We are left with the equation
\begin{widetext}
\begin{equation}\label{app:topophase}\tag{B.1}
\begin{pmatrix}
1+h^{\uparrow\uparrow}_k-\frac{1}{\alpha} & 0 & 0 & D^{\uparrow\uparrow}_k\\
0 & \frac{1}{\alpha}-1-h^{\uparrow\uparrow}_{k} & D^{\uparrow\uparrow}_{k} & 0\\
0 & D^{\uparrow\uparrow}_{k} & 1+h^{\uparrow\uparrow}_{k} + \frac{1}{\alpha} & 0 \\
D^{\uparrow\uparrow}_k & 0 & 0 & -1-h^{\uparrow\uparrow}_{k}-\frac{1}{\alpha}
\end{pmatrix}
\Psi_k = 0,\quad k = 0,\pm \pi.
\end{equation}
\end{widetext}
Taking the determinant of the matrix to be zero and solving for $\alpha$ then returns Eqn.~(\ref{topodiag4}). The calculation does not require fixing $\theta$ (in fact, the solution will not depend on $\theta$ at all), but the parameter curve thus obtained does not take gapless phases into account and is therefore of limited value unless $\theta = \pi/2$.

Calculating the determinant can be done in several ways, but we will employ the following identity, valid for block matrices where the upper diagonal block matrix is invertible:
\begin{equation}\tag{B.2}
\det\begin{pmatrix}
A & B\\ C & D
\end{pmatrix} = \det A\det(D-BA^{-1}C).
\end{equation}
Applying the above relation to eq.~(\ref{app:topophase}) gives us
\begin{equation} \label{app:topodet}\tag{B.3}
\det\left((1+h^{\uparrow\uparrow}_k-\frac{1}{\alpha})\sigma_z\right) = 0,
\end{equation}
or
\begin{equation}\tag{B.4}
\det\left((1+h^{\uparrow\uparrow}_k+\frac{1}{\alpha}+\frac{(D^{\uparrow\uparrow}_k)^2}{1+h^{\uparrow\uparrow}_k-\frac{1}{\alpha}})\sigma_z\right) = 0,
\end{equation}
where $\sigma_z$ is the diagonal Pauli-matrix. The upper equation is nothing but the non-linear two-band model, while the right equation gives us the new four-band phase diagram. Rewriting both of the equations leads us to 
\begin{equation}\tag{B.5}
\alpha(1+h^{\uparrow\uparrow}_k)=1,\ \text{and}\ \alpha^2((1+h^{\uparrow\uparrow}_k)^2 + (D^{\uparrow\uparrow}_k)^2) = 1,
\end{equation}
from which it is evident that the two equations are approximately equal when 
\[
(1+h^{\uparrow\uparrow}_k)^2 \gg (D^{\uparrow\uparrow}_k)^2,
\]
or equivalently
\begin{align}
(\tilde{f}(k_F+&k_H)+\tilde{f}(k_F-k_H))^2 \ll \nonumber\\
&4(\tilde{A}(k_F+k_H)+\tilde{A}(k_F-k_H)+k_Fa)^2,\tag{B.6}
\end{align}
where $\tilde{f}(k)$ and $\tilde{A}(k)$ are the same as $f(k)$ and $A(k)$ defined in the article, except for a sign change on the trigonometric terms if $k = \pm\pi$.
The above inequality is necessarily satisfied when $k_Fa \gg 1$, as both $\tilde{f}(k)$ and $\tilde{A}(k)$ are bounded for all finite $\xi_0$. It is important to note that the left-hand equation in~(\ref{app:topodet}) is the non-linear two-band model and does not correspond to a real root of the original determinant. This is because our approach becomes invalid, since right-hand equation becomes undefined because of the non-invertibility of our $A$ matrix. As a result of this, we see that the four-band model approaches the two-band model when $k_Fa$ grows and given that the non-linear two-band model reduces to the linear model when $\alpha\to 1$ we can conclude that our model reduces to the linear two-band model in the deep-dilute limit.

\section*{Appendix C: Analytical solution of the four-band model}
Beginning with Eq.~(\ref{ShibaANA}), we take the determinant of the complete matrix and require that it be zero. While this is at first glance a polynomial equation of degree 8, straightforward manipulation of terms will reduce it to an equation of the form
$
a\lambda^4 + b\lambda^3 + c\lambda^2 - b\lambda + a = 0
$, where all the parameters are functions of $k$. Because of the simple form of this fourth-degree polynomial equation, the solutions are reasonably short:
\begin{align}\label{app:fullsol}
&\lambda_{\beta\gamma}(k) = \beta \frac{\sqrt{b^2-4ac-8a^2}}{4a}+\nonumber\\
&+\frac{\gamma}{2}\sqrt{\frac{b^2}{2a} + \beta\frac{8b+4\frac{bc}{a} - (\frac{b}{a})^3}{2\sqrt{b^2-4ac-8a^2}} - \frac{c}{a} + 2} - \frac{b}{4a}\tag{C.1}
\end{align}
where $\beta = \pm 1$, $\gamma = \pm 1$. This assumes $a \neq 0$, which is the case for all parameters used in this article. The terms in the quartic equation defined above are found to be
\begin{widetext}
\begin{equation}\tag{C.2}
\begin{split}
a = &\alpha^2\left[(h_{k}^{\uparrow\downarrow})^2-(1+h_k^{\uparrow\uparrow})(1+h_{-k}^{\uparrow\uparrow})\right]\\
 b = &\alpha^3\left[(1+h_k^{\uparrow\uparrow})(D_{-k}^{\uparrow\uparrow})^2-(1+h_{-k}^{\uparrow\uparrow})(D_k^{\uparrow\uparrow})^2+2D_{k}^{\uparrow\downarrow}h_{k}^{\uparrow\downarrow}(D_k^{\uparrow\uparrow}-D_{-k}^{\uparrow\uparrow})\right]\\
& + \alpha^3(h_{-k}^{\uparrow\uparrow}-h_k^{\uparrow\uparrow})\left[(1+h_k^{\uparrow\uparrow})(1+h_{-k}^{\uparrow\uparrow})+(D_{k}^{\uparrow\downarrow})^2 -(h_{k}^{\uparrow\downarrow})^2+\alpha^{-2}\right]\\ 
c = &\alpha^4 \left[(D_{k}^{\uparrow\downarrow})^2-(h_{k}^{\uparrow\downarrow})^2+(1+h_k^{\uparrow\uparrow})(1+h_{-k}^{\uparrow\uparrow})-D_k^{\uparrow\uparrow}D_{-k}^{\uparrow\uparrow}\right]^2\\
&+\alpha^4\left[2D_{k}^{\uparrow\downarrow}h_{k}^{\uparrow\downarrow}-(1+h_k^{\uparrow\uparrow})D_{-k}^{\uparrow\uparrow}-D_k^{\uparrow\uparrow}(1+h_{-k}^{\uparrow\uparrow})\right]^2\\
 &+\alpha^2\left[2(D_{k}^{\uparrow\downarrow})^2-(D_k^{\uparrow\uparrow})^2-(D_{-k}^{\uparrow\uparrow})^2 - (h_k^{\uparrow\uparrow}-h_{-k}^{\uparrow\uparrow})^2\right]
+1
\end{split}
\end{equation}
\end{widetext}
In total we obtain four energy bands, 
\begin{equation}\label{app:energies}\tag{C.3}
E_{\beta\gamma}(k) = |\Delta|\frac{\lambda_{\beta\gamma}(k)^2-1}{\lambda_{\beta\gamma}(k)^2+1}
\end{equation}
with a high degree of symmetry: there is only one independent band, and the other three can be obtained by reflecting this over $E=0$ and/or $k=0$. Two of these bands are redundant solutions stemming from the mathematical formalism used; the correct bands are the ones that reduce to the two-band solutions in the deep-dilute limit (these turn out to be $E_{++}(k)$ and $E_{-+}(k)$).

If we further restrict the system to $\theta = \pi/2$, the term denoted $b$ vanishes, resulting in an equation of the form $a\lambda^4 + c\lambda^2 + a = 0$, from which it is easy to solve $\lambda^2$:
\begin{equation} \label{app:planarsol}\tag{C.4}
\lambda^2 = -\frac{c}{2a} \pm \sqrt{\frac{c^2}{4a^2} - 1}
\end{equation}
where
\begin{equation}\tag{C.4}
\begin{split}
a = &\alpha^2\left[(h_k^{\uparrow\downarrow})^2-(1+h_k^{\uparrow\uparrow})^2\right]\\
c = &\alpha^4 \left[(D_{k}^{\uparrow\downarrow})^2-(h_{k}^{\uparrow\downarrow})^2+(1+h_k^{\uparrow\uparrow})^2-(D_k^{\uparrow\uparrow})^2\right]^2\\
+&\alpha^4\left[\left(2D_{k}^{\uparrow\downarrow}h_{k}^{\uparrow\downarrow}-2(1+h_k^{\uparrow\uparrow})D_{k}^{\uparrow\uparrow}\right)^2-(D_{k}^{\uparrow\downarrow})^2(h_{k}^{\uparrow\downarrow})^2\right]\\
+&\alpha^2\left[2(D_{k}^{\uparrow\downarrow})^2-2(D_k^{\uparrow\uparrow})^2\right]+1\\
\end{split}
\end{equation}
Since the energy only depends on $\lambda^2$, we only find two separate energy bands, one with negative and one with positive energy and both of which are $k$-symmetric.
\vspace{20mm}
\begin{widetext}
 
\end{widetext}


\begin{thebibliography}{16}


\bibitem{yu} L. Yu, Acta Phys. Sin.  \textbf{21}, 75 (1965).
\bibitem{shiba} H. Shiba, Prog. Theor. Phys.  \textbf{40}, 435 (1968).
\bibitem{rusinov} A. I. Rusinov, JETP Lett.  \textbf{9}, 85 (1969).
\bibitem{salkola} M. I. Salkola, A. V. Balatsky and J. R. Schrieffer, Phys. Rev. B \textbf{55}, 12648 (1997).
\bibitem{yaz} A. Yazdani, B. A. Jones, C. P. Lutz, M. F. Crommie, and D. M. Eigler, Science \textbf{275}, 1767 (1997).
\bibitem{choy} T. P. Choy, J. M. Edge, A. R. Akhmerov, and C. W. J. Beenakker, Phys. Rev. B \textbf{84}, 195442 (2011).
\bibitem{np} S. Nadj-Perge, I. K. Drozdov, B. A. Bernevig, and A. Yazdani, Phys. Rev. B \textbf{88}, 020407(R) (2013).
\bibitem{klinovaja2}J. Klinovaja, M. J. Schmidt, B. Braunecker, and D. Loss, Phys. Rev. Lett. \textbf{106}, 156809 (2011).
\bibitem{klinovaja}J. Klinovaja, P. Stano, and D. Loss, Phys. Rev. Lett. \textbf{109}, 236801 (2012).
\bibitem{kjaergaard}M. Kjaergaard, K. W\"olms, and K. Flensberg, Phys. Rev. B \textbf{85}, 020503 (2012).
\bibitem{ojanen2}T. Ojanen, Phys. Rev. B \textbf{88}, 220502  (2013).
\bibitem{lutchyn}R. M. Lutchyn, J. D. Sau, and S. Das Sarma, Phys. Rev. Lett. \textbf{105}, 077001 (2010).
\bibitem{oreg}Y. Oreg, G. Refael, and F. von Oppen, Phys. Rev. Lett. \textbf{105}, 177002 (2010).
\bibitem{mourik}V. Mourik, K. Zuo, S. M. Frolov, S. R. Plissard, E. P. A. M. Bakkers, and L. P. Kouwenhoven, Science \textbf{336}, 6084 (2012).
\bibitem{das} A. Das, Y. Ronen, Y. Most, Y. Oreg, M. Heiblum, and H Shtrikman, Nat. Phys. \textbf{8}, 887 (2012).
\bibitem{np2}S. Nadj-Perge, I. K. Drozdov, J. Li, H. Chen, S. Jeon, J. Seo, A. H. MacDonald, B. Andrei Bernevig, and Ali Yazdani, Science 1259327 (published online 2 october 2014).
\bibitem{li}  J. Li, T. Neupert, B. A. Bernevig and  A. Yazdani, arXiv:1404.4058.
\bibitem{nayak} C. Nayak, S. H. Simon, A. Stern, M. Freedman, and S. Das Sarma, Rev. Mod. Phys. \textbf{80}, 1083 (2008).
\bibitem{vazifeh} M.M. Vazifeh, M. Franz, Phys. Rev. Lett. \textbf{111}, 206802 (2013).
\bibitem{poyh}K. P\"oyh\"onen, A. Weststr\"om, J. R\"ontynen and T. Ojanen, Phys. Rev. B \textbf{89}, 115109 (2014).
\bibitem{pientka2} F. Pientka, L. I. Glazman and F. von Oppen, Phys. Rev. B \textbf{88}, 155420 (2013).
\bibitem{pientka3} F. Pientka, L. I. Glazman and F. von Oppen, Phys. Rev. B \textbf{89}, 180505 (2014).
\bibitem{bry}  P. M. R. Brydon, H.-Y. Hui, J. D. Sau, arXiv:1407.6345. 
\bibitem{kitaev1} A. Y. Kitaev, Phys. Usp. \textbf{44}, 131 (2001).
\bibitem{heimes}  A. Heimes, P. Kotetes, G. Sch\"on, Phys. Rev. B \textbf{90}, 06050(R) (2014).
\bibitem{ront} J. R\"ontynen and T. Ojanen, arXiv:1406.4288.
\bibitem{klinovaja3}  J. Klinovaja, P. Stano, A. Yazdani and D. Loss, Phys. Rev. Lett. \textbf{111}, 186805 (2013).
\bibitem{braun}  B. Braunecker and P. Simon,  Phys. Rev. Lett. \textbf{111}, 147202 (2013).
\bibitem{reis} I. Reis, D. J. J. Marchand, and M. Franz, Phys. Rev. B \textbf{90}, 085124 (2014).

\end{thebibliography}
\end{document}